\documentstyle[12pt]{article}

\def\be{\begin{equation}}
\def\ee{\end{equation}}
\def\bear{\be\begin{array}}
\def\eear{\end{array}\ee}
\def\bea{\begin{eqnarray}}
\def\bea*{\begin{eqnarray*}}
\def\eea{\end{eqnarray}}

\def\draftlabel#1{{\@bsphack\if@filesw {\let\thepage\relax
   \xdef\@gtempa{\write\@auxout{\string
      \newlabel{#1}{{\@currentlabel}{\thepage}}}}}\@gtempa
   \if@nobreak \ifvmode\nobreak\fi\fi\fi\@esphack}
        \gdef\@eqnlabel{#1}}
\def\@eqnlabel{}
\def\@vacuum{}
\def\draftmarginnote#1{\marginpar{\raggedright\scriptsize\tt#1}}
\def\draft{\oddsidemargin 0.0truein
        \def\@oddfoot{\sl preliminary draft \hfil
        \rm\thepage\hfil\sl\today\quad\militarytime}
        \let\@evenfoot\@oddfoot \overfullrule 3pt
        \let\label=\draftlabel
        \let\marginnote=\draftmarginnote
   \def\@eqnnum{(\theequation)\rlap{\kern\marginparsep\tt\@eqnlabel}%
\global\let\@eqnlabel\@vacuum}  }

\begin{document}

\begin{titlepage}

\thispagestyle{empty}

\title{{\bf Fermions on the electroweak string}\thanks{Work partly supported
by CICYT under contract AEN94-0928, and by the European Union under
contract No. CHRX-CT92-0004}}

\author{ \\ {\bf J.M. Moreno},
{\bf D.~H. Oaknin}
\\  \\
Instituto de Estructura de la Materia, CSIC\\
Serrano 123, 28006-Madrid, Spain\\
and \\  \\
{\bf M. Quir\'os}\thanks{On leave from Instituto de Estructura de la
Materia, CSIC, Madrid, Spain}
\\ \\
Theory Division, CERN\\
CH-1211 Geneva 23, Switzerland  }

\date{}

\maketitle

\begin{abstract}
We construct a simple class of exact solutions of the
electroweak theory including the naked $Z$--string and
fermion fields. It consists
in the $Z$--string configuration
($\phi,Z_\theta$), the {\it time} and $z$
components of the neutral gauge bosons ($Z_{0,3},A_{0,3}$) and a
fermion condensate (lepton or quark) zero mode. The $Z$--string is not
altered (no feed back from the rest of
fields on the $Z$--string) while fermion
condensates are zero modes of the Dirac equation in the presence
of the $Z$--string background (no feed back from the {\it time}
and $z$ components of the neutral gauge bosons on the fermion fields).
For the case of the $n$--vortex
$Z$--string the number of zero modes found for charged leptons
and quarks is (according to previous results by Jackiw and
Rossi) equal to $|n|$, while for (massless) neutrinos is
$|n|-1$. The presence of fermion fields in its core make the
obtained configuration a superconducting string,
but their presence (as well as that of $Z_{0,3},A_{0,3}$) does not
enhance the stability of the $Z$--string.
\end{abstract}

\vspace{0.5cm}

\leftline{CERN-TH.7511/94}
\leftline{November 1994}

\vskip-21.0cm
\rightline{CERN-TH.7511/94}
\rightline{IEM-FT-95/94}
\vskip2cm

\thispagestyle{empty}
\end{titlepage}

%\end{document}

{\bf 1.} It is well known that cosmic strings are originated in
spontaneously broken gauge theories when the vacuum manifold is
not simply connected.
Strings originating at mass scales $\Lambda$
close to the Planck scale $M_{Pl}$ can yield (and be detected by)
gravitational effects: gravitational lensing, seeds for galaxy formation,
millisecond pulsar timing perturbations, etc. \cite{graveffects}.
On the other hand, strings originating at
$\Lambda \ll M_{Pl}$ have negligible gravitational effects and,
correspondingly,
cannot be detected through their gravitational interactions.
In particular,
the Nielsen-Olesen \cite{NO} vortex solution of the abelian Higgs model can
be embedded into the $SU(2)_L \times U(1)_Y$ electroweak theory
(and then $\Lambda  \sim G_F^{-1/2}$). This
vortex of $Z$-particles is known as the (naked) $Z$--string \cite{Zstring}.
As stated above $Z$--strings have negligible gravitational
interactions and experimental detection seems problematic,
though they have been proposed as candidates to
trigger baryogenesis at the electroweak phase transition
(no matter what the order is) \cite{baryo}.
However their dynamical stability (not
guaranteed by topological arguments) only holds for unrealistic
values of $\sin ^2 \theta_W$ \cite{stab}, a mechanism to stabilize the
$Z$--string being still
missing \footnote{See, however,\cite{Leandros} for a recent proposal.}.

Another class of cosmic strings with  non-gravitational effects
were proposed by Witten \cite{Witten1}.
They are called superconducting strings
because they have superconducting charge carriers
(either charged bosons or fermions)
with expectation values in the core of the string.
Superconducting strings can yield observable effects even for
$\Lambda \ll M_{Pl}$ ({\em e.g.} for
$ \Lambda \sim G_F^{-1/2}$) \cite{Witten2}.
In particular superconducting strings with Fermi charge
carriers can arise if there are normalizable fermion zero-modes
bounded to the string.

In this paper we prove that the $Z$--string is superconducting,
with leptons and quarks being the charge carriers. In
particular, we will embed the naked $Z$-string into a field
configuration with fermion fields, the {\em time} and $z$ component of the
electromagnetic, $A$, and $Z$ fields and the (unperturbed) $Z$--string.
Fermion condensates are zero modes of the Dirac
equation in the presence of the $Z$-string background. We have constructed
solutions where fermions are either charged leptons, quarks or neutral
leptons (neutrinos). As for the former (charged fermions) we have followed
the analysis of Ref. \cite{JR}. For the latter (neutrinos) we have
performed a similar analysis and found that neutrinos can be bounded
at the string core by the weak interactions. We have found that the
$Z$-string configuration is unaltered by the presence of both the fermion
condensate and the {\it time} and $z$ components of the electromagnetic
and $Z$ fields. A complete numerical analysis is also presented, including
the profiles for fermion densities and field configurations
and mean radii for the different bound states.

A similar analysis has been recently performed in
\cite{EP}, where a solution
of the electroweak theory with a single lepton family is constructed.
The authors of Ref.~\cite{EP} claim their solution is approximate since
they put $A_0=A_3=Z_0=Z_3=0$. However
we have explicitly shown that the fields $A_0$, $A_3$, $Z_0$ and $Z_3$
are non--zero in the presence of fermionic densities
and that {\it indeed} there is no feed
back from them on the fermion zero modes. It is also explicitly proven
in \cite{EP} that in the absence of the fields
$A_0$, $A_3$, $Z_0$ and $Z_3$ the presence of the fermion condensate
does not alter the stability properties of the $Z$-string. We have
proven by symmetry arguments that this is also the case when
dealing with the total solution containing non--zero values for
$A_0$, $A_3$, $Z_0$ and $Z_3$. We have also shown the above features
remain unchanged when zero--modes are either charged (leptons or quarks)
or neutral (neutrinos) fermions.

\vspace{1cm}

{\bf 2.} We will consider the case of just one fermion ({\it i.e.}
lepton or quark) species. Let
$\Psi = \left[ \begin{array}{c} \psi^+_L \\ \psi^-_L \end{array}\right]$
be the left--handed fermionic doublet and $\psi^+_R$, $\psi^-_R$ their
right--handed partners \footnote{The +/- superscript refers to the
up/down component, {\it i.e.} $\nu_{\ell}/\ell$ for leptons and
$u/d$ for quarks.
The absence of right--handed
neutrino in the Standard Model implies, in our notation,
$\psi^+_R$ = 0, $h_+ = 0$ for leptons.}.
The relevant lagrangian density is therefore:
\bear{ccl}
\label{lagrangian}
  {\cal L} & = &  -  \frac{1}{4} W_{\mu \nu \; a} W^{\mu \nu \; a}
                  -  \frac{1}{4} B_{\mu \nu}      B^{\mu \nu}
                  +  \left| D_{\lambda} \Phi \right|^2
                  -  \lambda (\Phi^{\dag} \Phi - \eta^2)^2  \\
           &   &  +  \; i {\bar \Psi} {\not \! \! D} \Psi \;
                  +     i {\bar \psi^+_R} {\not \! \! D}  \psi^+_R\;
                  +     i {\bar \psi^-_R} {\not \! \! D}  \psi^-_R\; \\
           &   &  -  h_+ \left( {\bar \Psi} {\tilde \Phi} \psi^+_R
                               + h.c. \right)
                  -  h_- \left( {\bar \Psi} \Phi \psi^-_R + h.c. \right)
\eear
 where
$W_{\mu \nu}^a = \partial_{\mu} W_{\nu}^a -
                 \partial_{\nu} W_{\mu}^a  +
                 g \varepsilon^{abc} W_{\mu}^b W_{\nu}^c$,
$B_{\mu \nu}  = \partial_{\mu} B_{\nu}     - \partial_{\nu} B_{\mu}$,
$ D_{\mu} = \partial_{\mu} + i g T^a W_{\mu}^a + i g' Y B_{\mu}$,
$T^a$ being the corresponding $SU(2)_L$  generator and $Y$ the
$U(1)_Y$
hypercharge.
$\Phi = \left[ \begin{array}{c} \phi^+ \\ \phi \end{array}\right]$
is the Higgs doublet and
${\tilde \Phi} \equiv i \sigma^2 \Phi^*$.

The Euler-Lagrange equations are:
\bear{ccl}
\label{eqnold}
D_\mu D^\mu \Phi       & = & - 2 \lambda (\Phi^{\dag} \Phi - \eta^2) \Phi
                           - h_+ \, ({\bar \Psi}\psi^+_R i \sigma^2)^t
                           - h_- \, {\bar \psi^-}_R \Psi \\
& & \\
(D_\nu W^{\mu \nu})^a  & = & -i \frac{g}{2} (\Phi^{\dag} \sigma^a (D^\mu \Phi)
 - (D^{\mu} \Phi)^{\dag} \sigma^a \Phi )
 - \frac{g}{2} \,{ \bar \Psi} \gamma^{\mu} \sigma^a \Psi \\
& & \\
 D_\nu B^{\mu \nu}     & = & -i \frac{g'}{2}(\Phi^{\dag}  (D^\mu \Phi)
 - (D^{\mu} \Phi)^{\dag} \Phi ) \\
& &
 -  g' y_L   \, \bar{\Psi}      \gamma^{\mu}  \Psi
 -  g' y_R^+ \, \bar{\psi_R^+}  \gamma^{\mu}  \psi_R^+
 -  g' y_R^- \, \bar{\psi_R^-}  \gamma^{\mu}  \psi_R^- \\
& & \\
i {\not \! \! D} \Psi      & = &  h_+ {\tilde \Phi} \psi^+_R +
                                  h_-         \Phi  \psi^-_R  \\
& & \\
i {\not \! \! D} \psi^+_R  & = &  h_+ {\tilde \Phi}^{\dag} \Psi \\
& & \\
i {\not \! \! D} \psi^-_R  & = &  h_- \Phi^{\dag} \Psi \\
\eear

To solve eqs. (\ref{eqnold})  it is necessary
to make an ansatz on the symmetry of the solution.
Our aim is to find some configuration that could be interpreted as
the $Z$-string  plus a fermion condensate in its core.
Then, our starting point will be to  generalize the
$Z$-string solution keeping, if possible, their symmetries.
In particular, concerning global symmetries, the $Z$-string
is invariant under the $Z_2$  {\em parity} given by the global
$U(2)$ transformation on the bosonic fields
$\left( \begin{array}{rc} -1 & 0 \\ 0 & 1 \end{array} \right)$.
The even fields under this {\em parity} are:
$\left[ \begin{array}{c}  0  \\ \phi \end{array} \right] $,
$W_\mu^3$, $B_\mu$.
There are several possible ways to extend this symmetry to
the fermionic fields. For example, we can choose one
fermion to be odd and the second one, that will be designed
by $\psi_L$, $\psi_R$, to be even \footnote{From here on, and for
notational simplicity, we will drop the $\pm$ superscript from
even fields.}.
It is clear from (\ref{eqnold}) that we can consistently fix to zero
all the odd fields.
Thus, the solutions under this ansatz are equivalent to those
in the reduced $U(1) \times U(1)$ model
     \footnote{Of course, the question of the stability should be
               addressed in the whole $SU(2) \times U(1)$ model.},
spontaneously broken to the $U(1)_{em}$ with $A_{\mu}$ and  $Z_{\mu}$
the corresponding gauge bosons.

The relevant field equations are obtained directly from (\ref{eqnold})
by replacing $\Phi \rightarrow \phi$, $\Psi \rightarrow \psi_L$;
$W^{1,2}, \phi^+ \rightarrow 0$. In terms of the mass
eigenstates gauge bosons,
$A_\mu = \sin \theta_W \, W^3_\mu + \cos \theta_W \, B_\mu$;
$Z_\mu = \cos \theta_W \, W^3_\mu - \sin \theta_W \, B_\mu$,
the covariant derivatives are:
\bear{ccl}
\label{covariant}
D_{\mu} \phi    & = &  ( \partial_\mu + i q_H Z_\mu              ) \; \phi \\
&& \\
D_{\mu} \psi_L  & = &  ( \partial_\mu + i q_L Z_\mu + i q A_\mu  ) \; \psi_L \\
&& \\
D_{\mu} \psi_R  & = &  ( \partial_\mu + i q_R Z_\mu + i q A_\mu  ) \; \psi_R
\eear
$q$ being the electric charge of the corresponding field and
$q_{H,\,L,\,R}$ the eigenvalues for the Higgs boson and left and
right fermions, respectively, of the $Z$-charge, defined in our
notation as
$Q^Z  = \frac{e}{\sin\theta_W \cos\theta_W}(T_3 - \sin^2 \theta_W \,q/e ) $,
where $T_3$ is the third component of the weak isospin, equal to 0
for singlets and  $\pm 1/2$ for the doublet components.

In particular, for the gauge bosons we have
\bear{ccl}
\label{eqAZ}
\Box Z^\mu - \partial^\mu \partial^\nu Z_\nu & = &
i q_H \left(  \phi^{\dag} (D^\mu \phi) - (D^{\mu} \phi)^{\dag} \phi  \right)
+ j_Z^{\mu} \\
&& \\
\Box A^\mu - \partial^\mu \partial^\nu A_\nu & = & j_A^{\mu}
\eear
where the fermionic currents on the right hand side of
(\ref{eqAZ}) are defined as
\bear{ccl}
\label{currents}
j_Z^{\mu} & = &  q_L \, {\bar \psi_L} \gamma^\mu \psi_L +
                         q_R \, {\bar \psi_R} \gamma^\mu \psi_R \\
&& \\
j_A^{\mu} & = &  q \, \left( {\bar \psi_L} \gamma^\mu \psi_L +
                              {\bar \psi_R} \gamma^\mu \psi_R \right) \\
\eear
as deduced from eq. (\ref{covariant}).

A general, static, $z$-independent ansatz still invariant under
the combined action of rotation around the z-axis and the suitable
gauge transformation is, in cylindrical coordinates \footnote{Here
$\alpha=0,3, \, r, \theta $ and we will write the equations in
the gauge $Z^r = A^r =0$}
\bear{lcl}
\label{bosons}
\phi  & = &  f(r) e^{- i n \theta} \\
Z^\alpha & = &  Z^\alpha(r)\\
A^\alpha & = &  A^\alpha(r)
\eear
for bosonic fields and
\bear{lcl}
\label{fermions}
\psi_L (r,\theta) & = &   \frac{1}{\sqrt{2}}\left[\begin{array}{l}
                      \psi_{L1}(r) \; e^{-im     \theta} \\
                      \psi_{L2}(r) \; e^{-i(m-1) \theta} \end{array}\right]
 \otimes \left[\begin{array}{r}  1   \\  -1  \end{array}\right]  \; \; \; \\
& &   \\
 \psi_R (r,\theta) & = &  \frac{1}{\sqrt{2}}\left[\begin{array}{l}
                    i \psi_{R1}(r) \; e^{-i(m \pm n)   \theta} \\
                    i \psi_{R2}(r) \; e^{-i(m \pm n
-1)\theta}\end{array}\right]
 \otimes \left[\begin{array}{r}  1   \\  1  \end{array}\right]
\eear
for
fermions \footnote{The $i$ factor in the parametrization
of $\psi_R$ is a matter of convention, and $\psi_{L1}(r)$,
$\psi_{L2}(r)$, $\psi_{R1}(r)$, $\psi_{R2}(r)$ are general complex
functions. Notice also that, since
we will look for zero modes, we have already fixed
the fermion energy $\omega$ to zero and no prefactor
$e^{-i\omega t}$ appears in (\ref{fermions}).},
where we are using the Dirac representation:
\be
\label{dirac}
\gamma^0 = \left( \begin{array}{cr} \sigma^0&     0      \\
                                        0  &  -\sigma^0 \end{array}\right) \;\;
\gamma^i = \left( \begin{array}{rc}     0  & \sigma^i  \\
                                - \sigma^i &   0       \end{array}\right) \;\;
\gamma^5 = \left( \begin{array}{cc}     0  & {\bf 1}   \\
                                 {\bf 1}   &  0        \end{array}\right)
\ee
with $\sigma^i$ the Pauli matrices and $\sigma^0 = {\bf 1}$.
Notice that an enlarged gauge configuration,
compared to the electroweak string,
is expected in general due to the presence of the fermionic
currents that will act as sources.

Using the ansatz (\ref{bosons},\ref{fermions}),
the $\theta$--dependence cancels in
the Euler--Lagrange equations. Defining
${\bf \psi_L}^t = (\psi_{L1}, \psi_{L2})$, ${\bf \psi_R}^t
= (\psi_{R1}, \psi_{R2})$
we get
\bear{rcl}
\label{ansatz}
f''    & = & - \frac{1}{r} f'
             + \left[ \frac{n^2}{r^2} - \frac{2n}{r}q_H Z_{\theta}
             - q_H^2 \left( Z_0^2 - Z_3^2  - Z_\theta^2 \right)
             + 2 \lambda(|f|^2 - \eta^2)  \right]  f \\
     &   &   \pm i h_\pm {\bf \psi_{L,R}}^{\dag} \sigma^0 {\bf \psi_{R,L}}
\eear
\bear{rcl}
\label{dos}
Z_0''      & = & -   \frac{1}{r} Z_0'
                 + 2 q_H^2 |f|^2 Z_0
                 - q_L \, {\bf \psi_L}^{\dag} \sigma^0 {\bf \psi_L}
                 - q_R \, {\bf \psi_R}^{\dag} \sigma^0 {\bf \psi_R} \\
Z_3''      & = & -   \frac{1}{r} Z_3'
                 + 2 q_H^2 |f|^2 Z_3
                 - q_L \, {\bf \psi_L}^{\dag} \sigma^3 {\bf \psi_L}
                 + q_R \, {\bf \psi_R}^{\dag} \sigma^3 {\bf \psi_R} \\
Z_\theta'' & = & -   \frac{1}{r} Z_\theta'
                 + 2 q_H^2 |f|^2 Z_\theta - 2 \frac{n}{r} q_H |f|^2
                 + \frac{1}{r^2} Z_\theta
                 - q_L \, {\bf  \psi_L}^{\dag} \sigma^2 {\bf \psi_L}
                 + q_R \, {\bf  \psi_R}^{\dag} \sigma^2 {\bf \psi_R} \\
0         & = & -2 q_H Im \left[ f^* f' \right]
                 + q_L \, {\bf  \psi_L}^{\dag} \sigma^1 {\bf \psi_L}
                 - q_R \, {\bf  \psi_R}^{\dag} \sigma^1 {\bf \psi_R}
\eear
\bear{rcl}
\label{tres}
A_0''      & = & -   \frac{1}{r} A_0'
                 - q \, {\bf \psi_L}^{\dag} \sigma^0 {\bf \psi_L}
                 - q \, {\bf \psi_R}^{\dag} \sigma^0 {\bf \psi_R} \\
A_3''      & = & -   \frac{1}{r} A_3'
                 - q \, {\bf \psi_L}^{\dag} \sigma^3 {\bf \psi_L}
                 + q \, {\bf \psi_R}^{\dag} \sigma^3 {\bf \psi_R} \\
A_\theta'' & = & -   \frac{1}{r} A_\theta'+ \frac{1}{r^2} A_\theta
                 - q \, {\bf \psi_L}^{\dag} \sigma^2 {\bf \psi_L}
                 + q \, {\bf \psi_R}^{\dag} \sigma^2 {\bf \psi_R}  \\
0        & = &     q \, {\bf  \psi_L}^{\dag} \sigma^1 {\bf \psi_L}
                 - q \, {\bf  \psi_R}^{\dag} \sigma^1 {\bf \psi_R}
\eear
where the last equations of (\ref{dos}) and (\ref{tres}) are
{\it constraints} corresponding to the gauge conditions $Z^r=A^r=0$, and
\bear{r c r l c r}
\sigma^1 {\bf \psi_L}' &  =  &
                            \{ \,\frac{1}{r} M_L  -  i q_L \sigma^2  Z_\theta
                       &  + \;i q_L \sigma^0 Z_0  - \, i q_L \sigma^3  Z_3
                       &    &
                             \\
                       &    &
                          - \;\;i  q  \sigma^2 A_\theta
                       &  + \;\,i  q  \sigma^0 A_0 \, -
                            \;\,i  q  \sigma^3 A_3  \; \} \,{\bf \psi_L}
                       &  - &   h \, f^\pm \sigma^0 {\bf \psi_R}  \\
&&&&& \\
\sigma^1 {\bf \psi_R}' &  =  &
                            \{ \,\frac{1}{r} M_R  -  i q_R \sigma^2  Z_\theta
                       &  - \;i q_R \sigma^0 Z_0  - \, i q_R \sigma^3  Z_3
                       &    &
                             \\
                       &    &
                          - \;\;i  q  \sigma^2 A_\theta
                       &  - \;\,i  q  \sigma^0 A_0 \, -
                            \;\,i  q  \sigma^3 A_3  \; \} \,{\bf \psi_R}
                       &  - &   h \, f^\pm \sigma^0 {\bf \psi_L}
\eear
where the prime denotes the derivative with respect to the cylindrical
radius $r$, and
\be
\label{angular}
 M_L  = \left( \begin{array}{cc}  0           &  m - 1 \\
                                  -m          &  0  \end{array}\right) \; \; \;
M_R  = \left( \begin{array}{cc}    0          &  m \pm n - 1 \\
                                  -(m \pm n)  &  0  \end{array}\right)
\ee
the $\pm$ sign corresponding to the case where the even
fermionic fields are $\psi_{L}^{\pm},\psi_{R}^{\pm}$ and
$f^+ = f^*$, $f^- = f$.

Before going ahead with our ansatz, we will review
some features of the vortex-fermion system.
Jackiw and Rossi showed in Ref. \cite{JR} that this system has
normalizable zero-modes if the fermions get their mass through their coupling
to the scalar field. In fact, the number of these zero modes depends on
the winding number of the vortex by an index theorem \cite{index}.
These zero modes are {\em transverse} \cite{Witten1},
{\it i.e.} they are eigenstates of  $\gamma^0 \gamma^3$,
\be
\label{eigen}
\gamma^0 \gamma^3 \psi = \kappa \, \psi, \;\; \kappa^2 =1\ ,
\ee
which is
the operator on the fermions associated to the parity operation
$(t,z) \rightarrow (-t,-z)$. Let us see how these zero
modes are in our case. Notice that, as we remarked before, the
gauge field configuration of the naked $Z$-string
must be enlarged in the presence of this fermionic density.
In particular, the {\em time} and $z$ components will be different from zero.

Suppose that (\ref{eigen}) also holds in our case and let us
see whether it is consistent with eqs. (\ref{ansatz}). Then
\be
j_A^{0} = \kappa \, j_A^{3}
\ee
and
\be
\label{a0a3}
A^0 = \kappa \, A^3
\ee
if the boundary conditions also obey this relation.
In the same  way,
\be
j_Z^{0} = \kappa  \, j_Z^{3}
\ee
then
\be
\frac{d^2}{d\,r^2} (Z^0 - \kappa \, Z^3)
 =  - \frac{1}{r} \frac{d}{d\,r} (Z^0 - \kappa \, Z^3)
    + 2 \, q_H |f|^2 (Z^0 - \kappa \, Z^3)
\ee
and again
\be
\label{nofeed}
Z^0 = \kappa \, Z^3
\ee
for appropriate boundary conditions.
Then,
\bear{lcr}
\label{nofeedback}
\left ( \gamma^0 Z_0 + \gamma^3 Z_3 \right) \psi_{L,R}(r,\theta) & = & 0 \\
\left ( \gamma^0 A_0 + \gamma^3 A_3 \right) \psi_{L,R}(r,\theta) & = & 0
\eear
where $\psi_{L,R}(r,\theta)$ are the spinors defined in the ansatz
(\ref{fermions}). Eqs. (\ref{nofeedback}) and (\ref{nofeed})
show that the fermionic zero
modes (\ref{eigen}) produce no feedback on the naked $Z$-string.
In other words, the ansatz (\ref{bosons})
is consistent with the existence of fermionic
zero modes since odd quantities under the operator $\gamma^0\gamma^3$
vanish.

Notice that the coefficients in the equations are real, and therefore the
phases of $f(r)$, ${\bf \psi_L}(r)$, ${\bf \psi_R}(r)$  are constant.
We can use the global $U(1) \times U(1)$ to bring them to zero.
It is easy to check that condition (\ref{eigen}) leads to the ansatz
\begin{equation}
\label{kappam1}
\psi_L=\left(
\begin{array}{c}
\psi_{L1} \\
   0
\end{array}
\right),
\ \
\psi_R=\left(
\begin{array}{c}
    0  \\
\psi_{R2}
\end{array}
\right)
\end{equation}
for $\kappa=-1$, and
\begin{equation}
\label{kappa1}
\psi_L=\left(
\begin{array}{c}
    0   \\
\psi_{L2}
\end{array}
\right),
\ \
\psi_R=\left(
\begin{array}{c}
\psi_{R1} \\
   0
\end{array}
\right)
\end{equation}
for $\kappa=1$. Using now the ansatz (\ref{kappam1}) or (\ref{kappa1}),
and $f(r)$ real, the constraints in (\ref{dos}) and (\ref{tres}),
corresponding to the gauge conditions $Z^r=A^r=0$, are trivially
satisfied.

The conclusions of this discussion can be summarized as follows:

\begin{itemize}

\item It is possible to choose consistently  fermions
(both left and right components)  in one $\kappa$-sector.
These configurations correspond to zero-modes.

\item In this case, the back reaction on the fields in the
      undressed electroweak string ($\phi$, $Z_\theta$) is
      {\em exactly} zero.
\item The equations for the fermions are just given by the Dirac
      equation in the electroweak string background.
\item The {\it time} and $z$ components of the gauge fields are
      different from zero.
\end{itemize}

As for the problem of the dynamical stability of this string
configuration, it holds for the same values of physical parameters
as in the case of the $Z$-string.
To see that, let us just consider, on top of the
ansatz (\ref{bosons}) and (\ref{fermions}), the same kind of
{\em dangerous} perturbations
$\delta \, W_{i}^{\pm}(r, \theta)$ ($i=1,2$),
$\delta \, \phi^{\pm}(r,\theta)$
that destabilize the $Z$-string.
These perturbations are {\em odd} in our language.
Let us focus, first, on fermionic terms.
As fermions always appear in cubic terms
with one boson, it is clear that this kind of perturbations decouple.
This was explicitly proven in Ref.~\cite{EP}.
It is also straightforward to see that these perturbations decouple
in the new bosonic terms induced in the energy by $Z_{0,3}$, $A_{0,3}$.
In fact the only relevant terms consistent with our ansatz and
Lorentz covariance are
\begin{eqnarray}
\label{WWZZ}
& &\delta  \, W_{i}^{\pm}(r, \theta)\delta \, W_{j}^{\mp}(r, \theta)
\delta^{ij}\times \\
& &\left\{ \alpha_W\left[A_0(r)^2-A_3(r)^2\right]+\beta_W
\left[Z_0(r)^2-Z_3(r)^2\right]
+\gamma_W\left[A_0(r)Z_0(r)-A_3(r)Z_3(r)\right]\right\} \nonumber
\end{eqnarray}
and
\begin{eqnarray}
\label{ppZZ}
&& \delta\, \phi^{\pm}(r,\theta)\delta \, \phi^{\mp}(r,\theta)\times \\
&&\left\{\alpha_\phi\left[A_0(r)^2-A_3(r)^2\right]+
\beta_\phi\left[Z_0(r)^2-Z_3(r)^2\right]
+\gamma_\phi\left[A_0(r)Z_0(r)-A_3(r)Z_3(r)\right]\right\} \nonumber
\end{eqnarray}
where the coefficients $\alpha_{W,\phi}$, $\beta_{W,\phi}$ and
$\gamma_{W,\phi}$ are some dimensionless combinations of the gauge
coupling constants. Using now the conditions (\ref{a0a3}) and
(\ref{nofeed}) one can easily check that the terms (\ref{WWZZ}) and
(\ref{ppZZ}) do vanish, as anticipated.
As we see, the requirement of simplicity that made this
configuration tractable is too strong to leave any room
for stability improvements.

%\newpage
\vspace{1cm}

{\bf 3.} We have found a consistent ansatz with zero energy,
but we still have to see if the equations admit a non--trivial,
normalizable solution. In the bosonic sector we take
$A_{\theta}(r)\equiv 0$, and, for the rest of the fields, the
boundary conditions
\begin{equation}
\begin{array}{c}
f(0)=0,\ f(\infty)=\eta  \\ \\
{\displaystyle
q_H Z_\theta\equiv \frac{v(r)}{r},\ v(0)=0,\ v(\infty)=-1 } \\ \\
A'_{0,3}(0)=Z'_{0,3}(0)=A'_{0,3}(\infty)=Z'_{0,3}(\infty)=0
\end{array}
\end{equation}
For the case of fermions getting their
masses through the coupling to the Higgs field
we follow the work of Jackiw and Rossi \cite{JR}.
The relevant equations are:
\begin{eqnarray}
\label{kplus}
\left[\begin{array}{l}  \psi_{L1}   \\
                        \psi_{R2}  \end{array}\right]' & = &
\left\{
\frac{1}{r}
\left(\begin{array}{cc } -m  & 0   \\
                          0  & m \pm n-1 \end{array}\right)
+ \; Z_\theta
\left( \begin{array}{cc}   q_L  &  0   \\
                           0    &  -q_R    \end{array}\right)
\right\}
\left[\begin{array}{l}  \psi_{L1}   \\
                        \psi_{R2}   \end{array}\right] \nonumber \\
& & \nonumber \\
& & -  h_{\pm} f
\left( \begin{array}{cc}     0  &  1   \\
                             1  &  0    \end{array}\right)
\left[\begin{array}{l} \psi_{L1}   \\
                       \psi_{R2}  \end{array}\right]
\end{eqnarray}

in the $\kappa=-1$ sector and

\begin{eqnarray}
\label{kminus}
\left[\begin{array}{l}  \psi_{L2}   \\
                        \psi_{R1}  \end{array}\right]' & = &
\left\{
\frac{1}{r}
\left(\begin{array}{ccc}  m - 1 &  & 0   \\
                          0     &  & -(m \pm n) \end{array}\right)
+ \;  Z_\theta
\left( \begin{array}{cc}   -q_L &    0   \\
                             0  &   q_R    \end{array}\right)
\right\}
\left[\begin{array}{l}  \psi_{L2}   \\
                        \psi_{R1}   \end{array}\right] \nonumber \\
& & \nonumber \\
& & -  h_{\pm} f
\left( \begin{array}{cc}     0  &  1   \\
                             1  &  0    \end{array}\right)
\left[\begin{array}{l} \psi_{L2}   \\
                       \psi_{R1}  \end{array}\right]
\end{eqnarray}

for $\kappa=1$.

The behaviour of $\psi_{L},\psi_{R}$ at {\em large} $r$ values
is controlled by the coupling to the Higgs field, because
the gauge field vanishes there.
In general, the asymptotic solution will be,
modulo some polynomial prefactors,
a combination of two exponential functions,
one increasing and another decreasing with $r$.
The requirement of normalizability translates into the condition
$\psi_L|_\infty =  \psi_R|_\infty $

For {\em small} $r$, we have again two equations coupled by the Higgs term
(the gauge term is negligible compared to the angular term for
wound fermions). By imposing consistency of the equations
and regularity of the spinors at the origin,
we get that just some values of $m$
(the parameter that controls the fermion winding)
are selected once $n$ is fixed. For these values,
the behaviour of the two fermion fields near $r=0$
is controlled by the {\it centrifugal} term coming from the angular
momentum, and the Higgs term is negligible.
The selected regions in the $n-m$ plane for the two $\kappa$ sectors and
the two $\psi^\pm$ fermions are shown in  Fig.~1.
Notice that the specific values of the fermionic gauge couplings
are irrelevant in the above considerations on the number of
zero modes. In fact, for the $n$-string there are $|n|$ zero-modes.

A quick glance at Fig.~1 shows that if we
try to put both $\psi^\pm$  fermions
in the same $\kappa$ sector
\footnote{In this case also $W^{\pm \, 0,3}$
is generated,
playing the same {\em r\^ole} as $(A,Z)^{0,3}$,
and the ansatz (\ref{bosons}) would be trivially enlarged.},
the solution cannot be normalized.
Therefore,
if we are looking for configurations involving both $\kappa$ sectors,
such as those with fermion energy different from zero,
the bosonic field ansatz must be enlarged.
Of course, it would be very interesting to find an exact
normalizable solution also in
this case, but the lost of the symmetry makes that task almost impossible.

So far, we have studied the normalizability of the solution
just for massive fermions. The {\em r\^ole} of the Higgs boson
was crucial there, and we can expect a very different situation
for massless fermions (neutrinos). The interaction of
the neutrinos with the $Z$-string is purely gauge and is governed by:
\be
\label{neutrino1}
\psi^{'}_{L1} =  \left( -\frac{m}{r} +
                        q_\nu Z_\theta \right) \psi_{L1}
\ee
for $\kappa = -1$ and
\be
\psi^{'}_{L2} =  \left( \frac{m-1}{r} -
                        q_\nu Z_\theta \right) \psi_{L2}
\ee
when  $\kappa =  1$. The behaviour of the neutrino field near
$r = 0$ is controlled by the angular term, and if we impose
$\psi$ to be regular at the origin we get:

\begin{center}
\begin{tabular}{rccccr}
 $- m$   &  $\geq$  & 0 &  \phantom{kkkk} for $\kappa$  & = &  $-1$  \\
 $m-1$   &  $\geq$  & 0 &  \phantom{kkkk} for $\kappa$  & = &  $ 1$
\end{tabular}
\end{center}

Due to the absence of the Higgs term, $\psi$ is not an exponential
function of $r$ at large values, but goes like $r^\alpha$, where
$\alpha$ is some integer.
In particular, for
$\kappa = -1$, $\psi_{L1} \sim r^{-(m+n)}$.
Then, for $(m+n) > 1$,
the neutrino distribution is normalizable.
If we ask for a localized configuration,
{\em i.e.} finite $\langle r \rangle$, $\langle r^2 \rangle$
and assume $m$ integer, {\it i.e.} periodic fermions
when winding around the string, we get:
\begin{center}
\begin{tabular}{rcrccr}
 $m+n$   &  $>$  &  $2$ &  \phantom{kkkk} for $\kappa$  & = &  $-1$  \\
 $m+n$   &  $<$  &  $-1$ &  \phantom{kkkk} for $\kappa$  & = &  $ 1$
\end{tabular}
\end{center}

We have found that there are normalizable fermionic
configurations with zero energy even in the
case of  massless particles. It means that
the gauge interaction can be efficient enough
to keep the neutrino near the string core.
Let us illustrate this situation by using an analogous
quantum mechanical system. From Eq. (\ref{neutrino1})
it follows:
\be
\left( - \frac{1}{2} \frac{d^2 }{dr^2} +  V_{eff} \right) \psi_{L1} = 0
\ee
where (for $\kappa$=--1)
\be
\label{veff}
 V_{eff}(r) = \frac{1}{2}\left\{   \frac{m(m+1)}{r^2} +
             q_\nu \frac{dZ_{\theta}(r)}{dr}
            + q_\nu^2 Z_{\theta}^2(r) - 2 m q_\nu
            \frac{Z_\theta(r)}{r} \right\}
\ee

This is the Schr\"odinger equation for a particle of unit mass
under the action of the potential described by $V_{eff}$.
In Fig.~2 we have drawn the  shape of this effective
potential as a function of the distance \footnote{All the dimensional
values  are expressed in units of the corresponding power of
$\langle \phi \rangle$.} to the string axis.
We have worked out two cases: one for $n=2$, $m=0$ ({\em i.e.}, without
centrifugal term) and the second one for $n=4$, $m=-2$
(with a centrifugal term).
Notice the existence of a centrifugal barrier when
$m \neq -1,0$. For large r values, $V_{eff} \sim r^{/2}$. For the
relevant values of $n$ and $m$, the structure of the potential
can be described by the existence of an absolute minimum
with $V_{min} < 0$ and a barrier with $V_{max} > 0$ that decreases as
$1/r^2$ for large $r$. As $V_{eff} (\infty) = 0$, we could
consider the possibility of tunnelling of the neutrino zero mode
to the large $r$ region, thus dissociating the bound state.
To estimate this probability,
we can use the WKB approximation \cite{Coleman} assuming that the potential
vanishes for $r>R$, with $R$ large enough, and take the
limit $R\rightarrow\infty$. It is easy to
see that this probability is
$\propto R^{-2\sqrt{(n+m)(n+m+1)}}$ and then goes to
zero \footnote{In a realistic situation, a natural value for $R$
could be provided by the radius of the string loop,
or by the typical distance between strings.}. We have also drawn in
the same picture the effective potential for a case in which
the distribution is normalizable but  $\langle r \rangle$ is
infinite.

\vspace{1cm}

{\bf 4.} We have solved numerically the field equations for the neutrino
and for massive fermions from the third generation. In Fig.~3a
we have plotted fermionic density per unit of string length
for several cases. They include the neutrino configuration for
$n=2$, $m=0$ and a fermion with $M=M_{top}$ in the limit
$q_L, q_R  \rightarrow 0$. For the neutrino, the interaction is
purely gauge, whereas for the other illustrating
fermion the interaction is only through the Higgs coupling.
We have also drawn the fermionic density for the top quark with the
same $n, m$ values. Notice that, as one could expect, the density
shape for the top quark almost coincides with the corresponding
one in the limit of zero charges.
In all cases, the typical mean radius is of the order of the
string radius. We have also calculated the $Z_0$ field component
corresponding to these configurations. The results are gathered in
Fig.~3b.
We have repeated this analysis for the bottom quark and the tau lepton.
The corresponding profiles are shown in Fig.~4a,b. Notice that the
density function goes to zero for large $r$ much more slowly than in
the case of the top quark. This is because the behaviour of the
wave function for massive fermions, as we said before, is described
by $e^{-M_f}r$, with $M_f$ the fermion mass. In Table 1 we list the
mean  and root-mean-squared radii corresponding to the configurations
shown in Figs.~3a,4a.
Finally, in Fig.~5a,b we show the radial electric field produced by the
charged fermionic distributions. The asymptotic form of this field is
given by the Gauss law, $E_r \sim q r^{-1}$ with $q$ the charge of the
fermion.

\vspace{1cm}

{\bf 5.} In summary, we have constructed in this paper a simple class
of exact solutions of the electroweak theory, which consists in the
$Z$-string configuration, the gauge fields $A_0$, $A_3$, $Z_0$ and $Z_3$,
and fermion zero modes bounded at the string core. We have explicitly
worked out the cases where fermions are charged leptons or quarks, and
neutral leptons (neutrinos). In the cases where zero modes are charged
fermions, moving under the action of the electromagnetic field, the
$Z$-string becomes superconducting. In our solution there is no feed back
from gauge and fermion fields on the $Z$-string configuration.
As for the stability problem our solution
does not add anything new with respect to the case of the $Z$-string
in the absence of fermion zero modes: either the $Z$-string configuration
is extended to a stable configuration including other bosonic fields, or
its stability is topologically guaranteed by
the presence of an extra spontaneously broken (gauge) symmetry,
remnant of the theory at some high scale \cite{frustrated}.
In both cases the presence of
fermion zero modes would make the string configuration superconducting
and therefore detectable by non-gravitational interactions.

%\newpage

\newpage

\begin{table}
\begin{center}
\begin{tabular}{|l|c|c|r|r|} \hline
&&&&\\
Particle   &   n   &  m  & $\langle r \rangle$ &
$\sqrt{\langle r^2 \rangle}$ \\
&&&&\\
\hline
neutrino   &   2     &  0  &  4.45  & -----  \\
top        &   1     &  0  &  1.18  &  1.37  \\
bottom     &  $-1$   &  0  & 16.39  & 23.12  \\
tau        &  $-1$   &  0  & 50.04  & 69.57  \\
\hline
\end{tabular}
%

%\vspace{1cm}
%Table 1
\caption{Mean and root-mean-square radii for fermionic configurations
shown in Fig.~4a,b}
\end{center}
\end{table}

\section*{\bf Figure captions}

\begin{description}
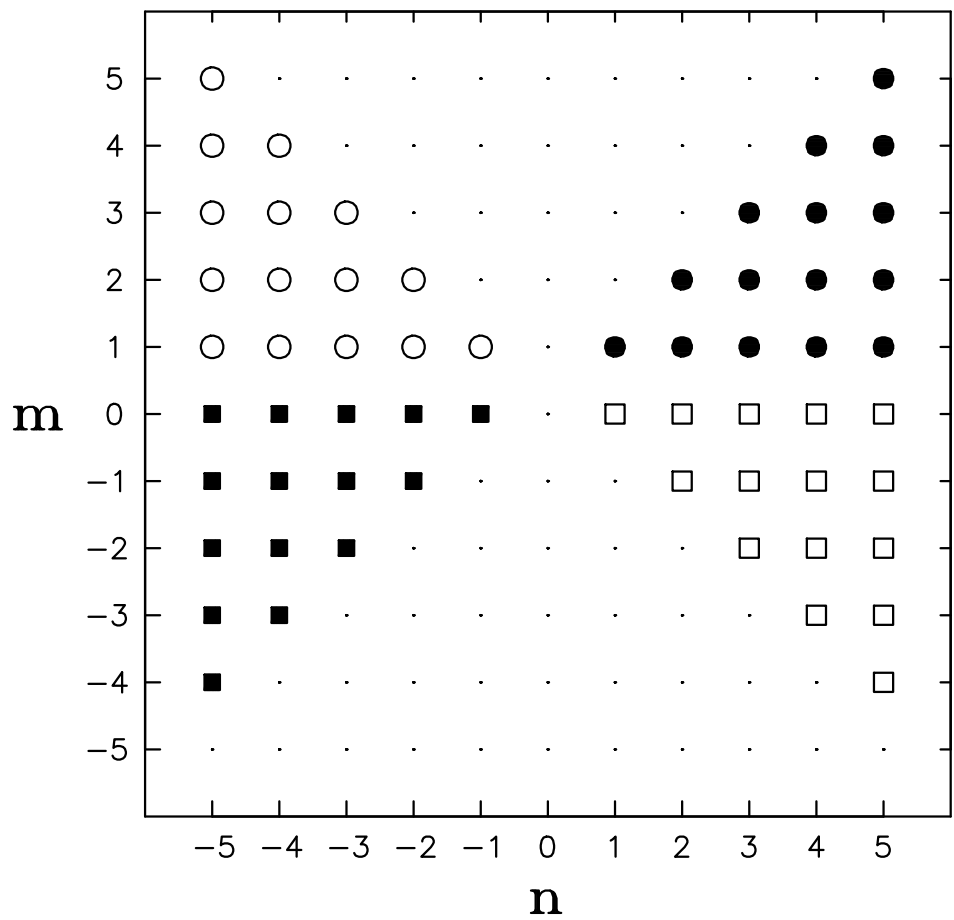
\item[{\bf Fig. 1}]
Allowed regions in the $(n,m)$ plane for $\kappa=1$ ($\kappa=-1$),
circles (squares), and $\psi^+$ ($\psi^-$), open (black), fermions.

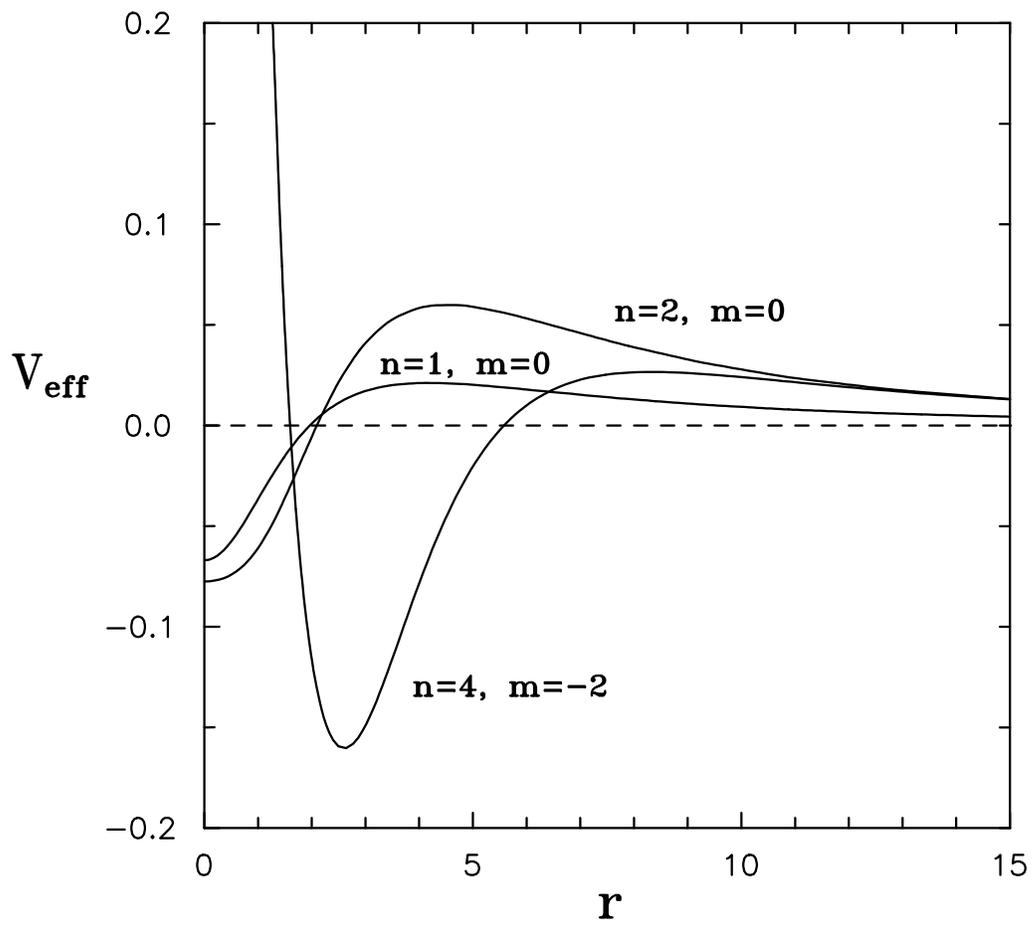
\item[{\bf Fig. 2}]
Plots of the effective potential
for the $\psi_{L1}$ neutrino component,
as defined in Eq. (\ref{veff}), for the indicated values of $(n,m)$.

\item[{\bf Fig. 3}]
{\bf a)} Fermionic densities for the top zero mode
$n=1, m=0$ and for the neutrino
configuration with $n=2, m=0$. For illustration, the density for
a massive top like fermion in the limit $q_{L,R} = 0$ is also
shown (dashed line). In all cases, $\kappa = -1$.
{\bf b)} $Z_0$ component of the gauge field generated by the
configurations shown in (a).

\item[{\bf Fig. 4}]
{\bf a)} Fermionic densities for tau and bottom fermionic zero modes.
In both cases, $n=-1, m=0$ and $\kappa = -1$.
{\bf b)} $Z_0$ component of the gauge field generated by the
configurations shown in (a).

\item[{\bf Fig. 5}]
{\bf a)} Electric field $E_r$ generated by the top quark configuration
shown in Fig. 3a.
{\bf b)} Electric field $E_r$ generated by the fermionic configurations
shown in Fig. 4a.

\end{description}

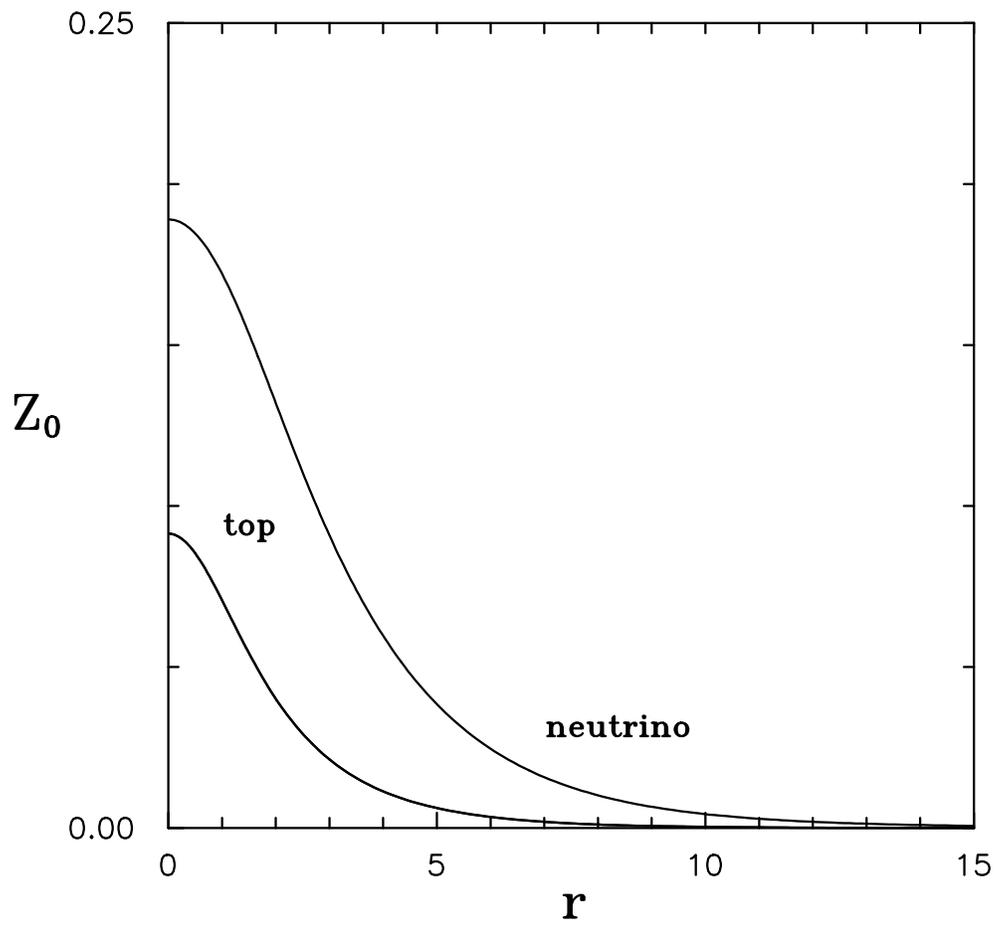
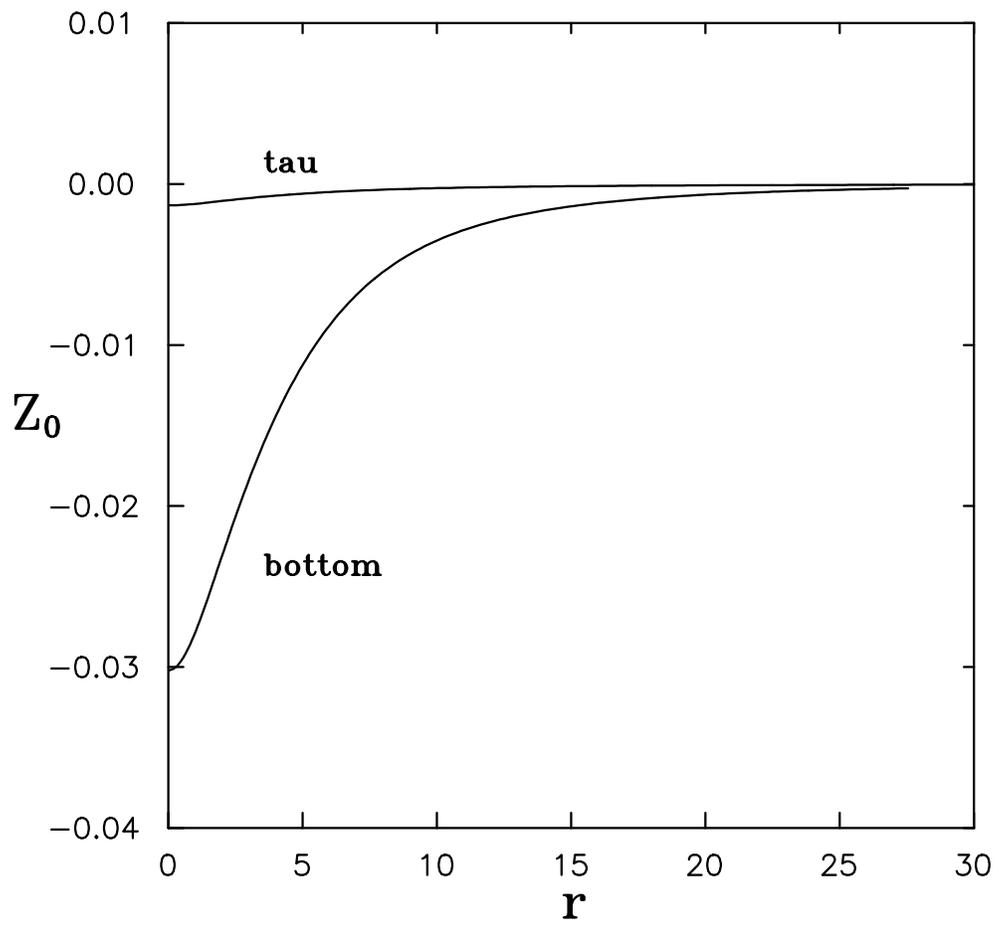
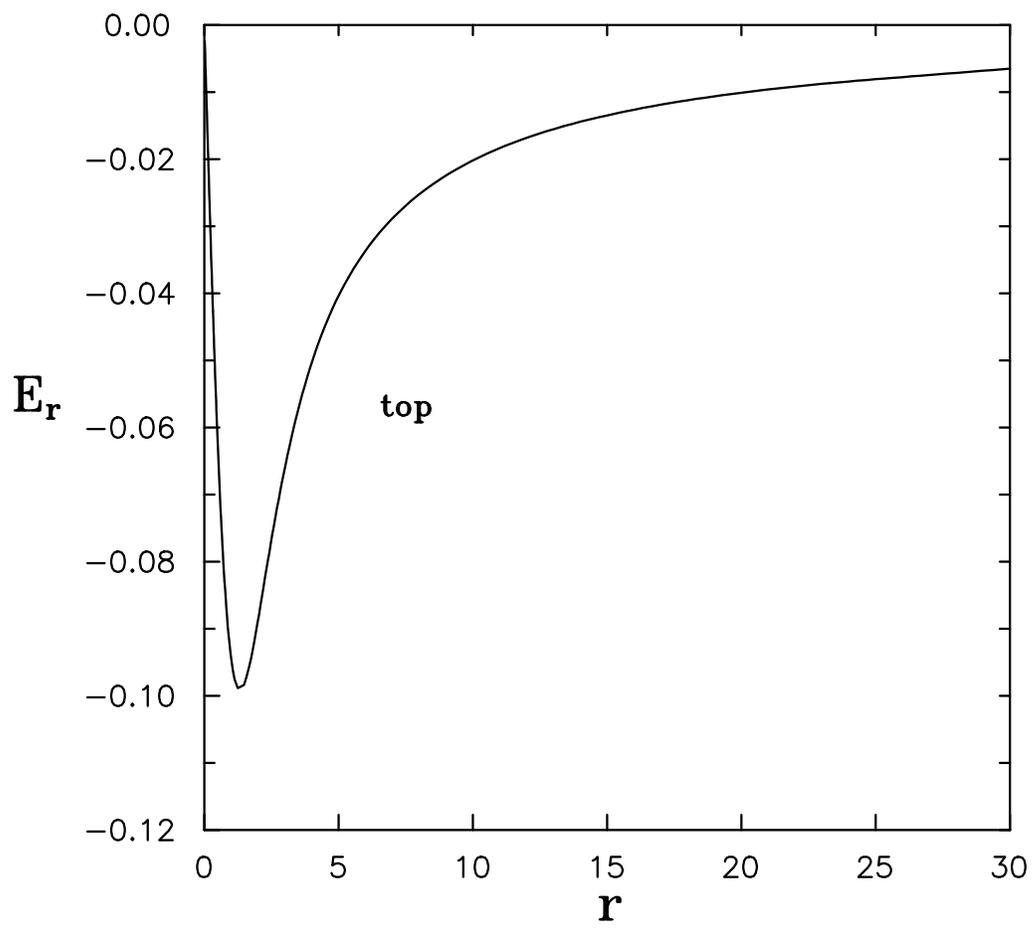
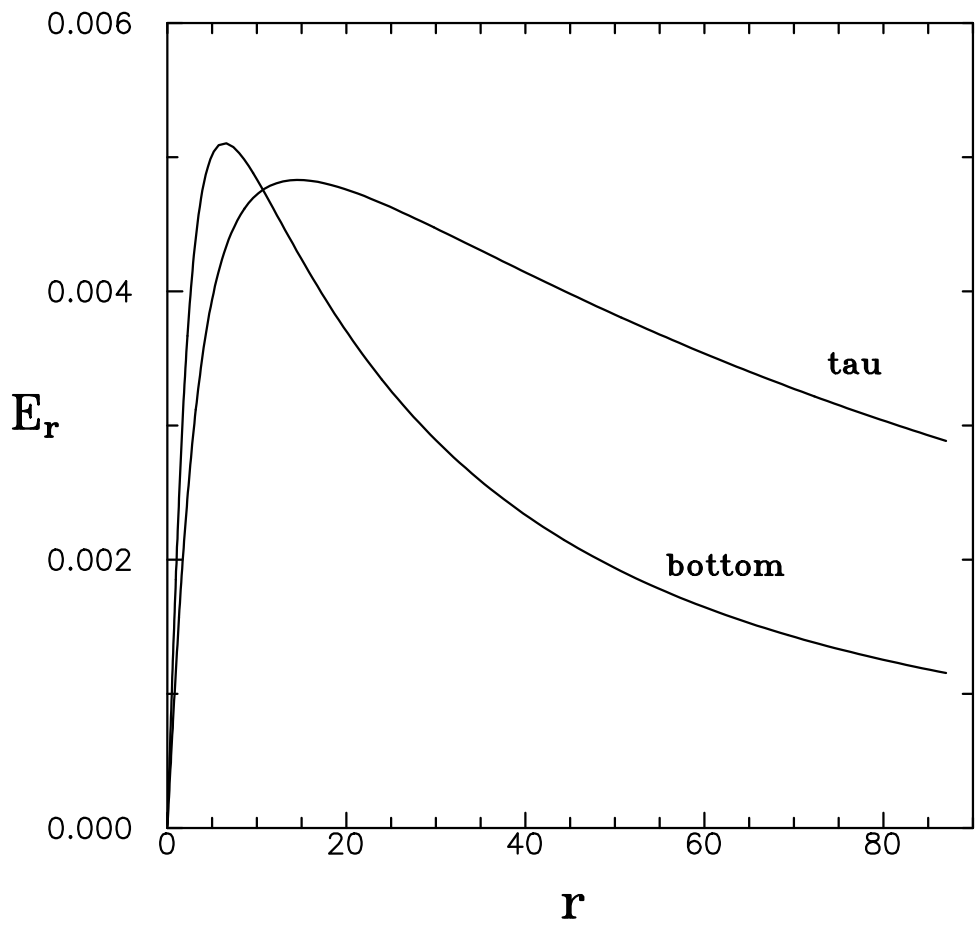
\end{document}